\documentclass{ws-procs9x6}            

\usepackage{color}
\usepackage[utf8]{inputenc}

\makeatletter
\newcommand\xleftrightarrow[2][]{%
  \ext@arrow 9999{\longleftrightarrowfill@}{#1}{#2}}
\newcommand\longleftrightarrowfill@{%
  \arrowfill@\leftarrow\relbar\rightarrow}
\makeatother

\newcommand{\hcM}{{\widehat{\mathcal{M}}}}
\newcommand{\hM}{{\widehat{M}}}
\newcommand{\hj}{{\widehat{j}}}
\newcommand{\tT}{{\tilde{T}}}
\newcommand{\wM}{{\widehat{M}}}
\newcommand{\tlambda}{{\tilde{\lambda}}}
\newcommand{\CM}{{\mathcal{C}^{\infty}(\mathcal{M})}}
\newcommand{\ChM}{{\mathcal{C}^{\infty}(\widehat{\mathcal{M}})}}
\newcommand{\cM}{{\mathcal{M}}}
\newcommand{\tM}{{\tilde{M}}}
\newcommand{\tq}{{\tilde{q}}}
\newcommand{\tp}{{\tilde{p}}}
\newcommand{\txi}{{\tilde{\xi}}}
\newcommand{\tx}{{\tilde{x}}}
\newcommand{\tB}{{\tilde{B}}}
\newcommand{\tbeta}{{\tilde{\beta}}}
\newcommand{\tV}{{\tilde{V}}}

\newcommand{\te}{{\tilde{e}}}

\newcommand{\tpar}{{\tilde{\partial}}}
\newcommand{\Sp}{{\;\,}}

\begin{document}
	
\markboth{M.A. Heller, N. Ikeda, S. Watamura}{Courant algebroids from double field theory in supergeometry}
\wstoc{Courant algebroids from double field theory in supergeometry}{M.A. Heller, N. Ikeda, S. Watamura}

\begin{flushright}
	\null \hfill Preprint TU-1040
\end{flushright}
	
\title{Courant algebroids from double field theory in supergeometry}

\author{Marc Andre Heller$^\dagger$, Noriaki Ikeda$^\ddag$ and Satoshi Watamura$^\sharp$}

\address{Particle Theory and Cosmology Group, \\
Department of Physics, Graduate School of Science, \\
Tohoku University \\
Aoba-ku, Sendai 980-8578, Japan \\
$^\dagger$E-mail: heller@tuhep.phys.tohoku.ac.jp \\
$^\sharp$E-mail: watamura@tuhep.phys.tohoku.ac.jp}

\address{$^\ddag$Department of Mathematical Sciences,
Ritsumeikan University \\
Kusatsu, Shiga 525-8577, Japan \\
E-mail: nikeda@se.ritsumei.ac.jp}

\begin{abstract}
We provide a short review of Courant algebroids and graded symplectic ma\-nifolds and show how different Courant algebroids emerge from double field theory. Fluxes of double field theory and their Bianchi identities are formulated by using pre-QP-manifolds. We stress the relation of the Poisson Courant algebroid with $R$-flux as a solution of the double field theory section condition.
\end{abstract}


\bodymatter
\vfill
\pagebreak

\section{Introduction}
\noindent

One of the most intriguing symmetries of string theory is T-duality. It is a target space symmetry relating backgrounds of compactified closed string theory. The appearance of this duality is due to the fact that strings as one-dimensional objects experience geometry differently compared to zero-dimensional particles. Especially closed strings can wrap internal cycles of a compactification manifold. The winding number then counts how often a closed string wraps a distinct cycle. On the other hand, the momentum of the string in compact directions is quantized. T-duality exchanges the winding modes and momentum modes, while inverting the scale of the compactification. 

The NS sector of string theory contains the so-called $3$-form $H$-flux, which can wrap internal cycles of the compactification. T-duality on backgrounds with Killing isometry was first investigated by Buscher in the '80s\cite{Buscher1, Buscher2}. In this case, T-duality mixes metric and $B$-field. A mathematical formulation making this symmetry manifest is generalized geometry\cite{Gualtieri:2003dx,Cavalcanti:2011wu}. It is well known, that several T-duality transformations of a $3$-torus with $H$-flux lead to so-called non-geometric backgrounds\cite{Shelton:2005cf,Wecht:2007ngf}. This situation is very well captured by the T-duality chain
\begin{equation}
	H_{abc} \xleftrightarrow{T_a} f^a_{bc} \xleftrightarrow{T_b} Q^{ab}_c \xleftrightarrow{T_c} R^{abc}, \notag
\end{equation}
where $T_a$ denotes T-duality in $x^a$-direction. $H$-flux is the field strength of the $2$-form $B$-field to which the string couples. The next step in the chain is the geometric $f$-flux which is related to the Weitzenb\"ock connection of the compactification manifold. The next step is the globally non-geometric $Q$-flux. The associated "manifold" is a so-called T-fold, which exhibits monodromies, that have to be patched by full T-duality transformations. The final step in the chain is the locally non-geometric $R$-flux. In this case, the ordinary manifold description breaks down, since even locally full T-duality transformations are necessary to patch the charts.

There have been many investigations of the ominous non-geometric backgrounds and associated non-geometric fluxes: The geometry of non-geometric backgrounds was analyzed in a paper by Hull\cite{Hull:2004in}. Non-geometric backgrounds and T-duality were also investigated from the viewpoint of generalized geometry\cite{Grana:2008yw,Blumenhagen:2012Bi}, topology\cite{Bouwknegt:2003vb,Bouwknegt:2004tr,Bouwknegt:2010zz} and non-associativity\cite{Blumenhagen:2011ph,Blumenhagen:2010hj,Mylonas:2012pg}. Non-geometric fluxes have also been shown to be related to so-called \emph{exotic branes}\cite{deBoer:2012ma,Hassler:2013wsa}. Dynamic fluxes in double field theory in relation with generalized Bianchi identities and the resulting differential geometry has also been investigated\cite{Geissbuhler:2013ed}.

In this article, we also want to discuss the \emph{Poisson Courant algebroid}, which exhibits a natural $3$-vector freedom\cite{Asakawa:2014kua,Bessho:2015tkk}. It is a Courant algebroid on a Poisson manifold and exchanges the roles of tangent and cotangent space compared to the Courant algebroid on the generalized tangent bundle which is closely related to generalized geometry. In the same manner, the Poisson Courant algebroid is closely related to what is called Poisson-generalized geometry\cite{Asakawa:2015aia}.

A T-duality invariant formulation of the effective string theory is double field theory\cite{Hull:2009mi,Aldazabal:2011nj}. Early developments of the ideas behind this theory reach back until the '90s\cite{Siegel:1993tv,Siegel:1993sd,Siegel:1993md}. Double field theory is a manifestly $O(D,D)$ invariant field theory. It is formulated on the double spacetime, which is parametrized by a double set of coordinates, standard coordinates $x^i$ and their duals $\tx_i$. Here, $i$ runs from $1$ to $D$, where $D$ is the dimension of the spacetime. The dual coordinates are associated to the winding modes of closed strings wrapping the internal cycles. The generators of $O(D,D)$ are $B$-transformations and diffeomorphisms, which span the geometric subgroup, as well as $\beta$-transformations. $\beta$-transformations are related to non-geometric orbits under T-duality, which corresponds to $O(D,D;\mathbb{Z})$-transformation after compactification of the theory on a $D$-dimensional torus. A supergravity formulation taking the $\beta$-potential of the $R$-flux as a starting point is the so-called $\beta$-supergravity\cite{Andriot:2013xca}. There have also been several investigations of non-geometric fluxes from the viewpoint of double field theory\cite{Andriot:2012an,Hassler:2014cc}.

In this article we review the formulation of various Courant algebroids as well as the Poisson Courant algebroid in terms of QP-manifolds, which are graded symplectic manifolds. Then, we discuss the formulation of double field theory using so-called pre-QP-manifolds\cite{DeserSaemann:2016,Heller:2016u}, a weaker version of QP-manifolds. The double field theory algebra has also been investigated in the context of graded symplectic manifolds\cite{Deser:2014mxa,Deser:2014ap,DeserSaemann:2016}. Finally, we interpret the Poisson Courant algebroid as a solution of the so-called \emph{strong constraint} or \emph{section condition} of double field theory and discuss the $3$-vector freedom on this basis\cite{Heller:2016u}.

This article is organized as follows. In section 2, we give a brief introduction in QP-manifolds and Courant algebroids accompanied with some simple examples. In section 3, a short review of double field theory and its relation to graded symplectic manifolds is presented. In section 4, we show how the local descriptions of geometric and non-geometric fluxes emerge by twist of a graded symplectic manifold and how this induces different Courant algebroids. Section 5 contains a review of the Poisson Courant algebroid as a Courant algebroid and from the viewpoint of QP-manifolds. In section 6, we discuss the Poisson Courant algebroid as a concept to model $R$-flux backgrounds and its relation to double field theory. Finally, section 7 is devoted to a summary and discussion.

\section{QP-manifolds and Courant algebroids}

In this section, we provide a brief introduction to the necessary means to understand the main text. First, we define the mathematical structure of a QP-manifold. Then, the structure of a Courant algebroid is introduced and related to QP-manifolds of degree $2$. Finally, we provide some examples of twisted Courant algebroids associated with twisted QP-manifolds. For a rigorous discussion of this matter we refer to Ref.~\citenum{Ikeda:2012laksz}.

\subsection{QP-manifolds}

Let us gather the elements that we need to define a QP-manifold. A \emph{graded manifold} can be locally described using graded coordinates. The grading is Grassmann even or odd. Then,  a non-negatively graded manifold $\cM$ is called an \emph{N-manifold}. Such an N-manifold can be equipped with a graded symplectic structure $\omega$ of degree $n$. Then, the pair ($\cM$, $\omega$) is called \emph{P-manifold} and $\omega$ is referred to as \emph{P-structure}. The graded symplectic structure induces a graded Poisson bracket, $\{-,-\}$, on the smooth functions over $\cM$ via
\begin{equation}
	\{f,g\} \equiv (-1)^{|f|+n+1}\iota_{X_f}\iota_{X_g}\omega,
\end{equation}
where $f,g\in\CM$ and $X_f$ is the associated Hamiltonian vector field to $f$ defined by
\begin{equation}
	\iota_{X_f} = - \delta f.
\end{equation}
The operator $\delta$ denotes the de Rham differential on the space of forms over $\cM$.

Let ($\cM$, $\omega$) be a P-manifold of degree $n$. A vector field $Q$ of degree $1$ over $\cM$, which is homological, $Q^2 = 0$, is called \emph{Q-structure}. Finally, a \emph{QP-manifold} ($\cM$, $\omega$, $Q$) is a P-manifold ($\cM$, $\omega$) together with a Q-structure $Q$, such that $L_Q \omega = 0$.

For any QP-manifold there exists a function $\Theta\in\CM$ of degree $n+1$, such that
\begin{equation}
	Qf = \{\Theta,f\},
\end{equation}
where $f\in\CM$. The function $\Theta$ is called \emph{Hamiltonian function}. Then, the nilpotency of the homological function, $Q^2 = 0$, translates to the \emph{classical master equation},
\begin{equation}
	\{\Theta, \Theta\} = 0.
\end{equation} 

Finally, we can define the operation of a \emph{twist} on a QP-manifold ($\cM$, $\omega$, $Q$) of degree $n$. Let $\alpha\in\CM$ a function of degree $n$. Then, the twist of any function $f\in\CM$ by $\alpha$ is defined using exponential adjoint action,
\begin{equation}
	e^{\delta_\alpha}f = f + \{f,\alpha\} + \frac{1}{2}\{\{f,\alpha\},\alpha\} + \cdots.
\end{equation}
Since the Poisson bracket is of degree $-n$, the operation is degree-preserving and we have
\begin{equation}
	\{e^{\delta_\alpha}f, e^{\delta_\alpha}g\} = e^{\delta_\alpha}\{f, g\},
\end{equation}
where $f,g\in\CM$.

\subsection{Courant algebroids}

Let us start by giving the definition of a Courant algebroid. Then, we will relate Courant algebroids to QP-manifolds of degree two and discuss some simple examples. Finally, we will comment on the operation of twisting in this setting.

Let $E$ be a vector bundle over a smooth manifold $M$. A \emph{Courant algebroid} consists of this vector bundle together with a bilinear operation $[-,-]_D$ on the sections of $E$, a bundle map $\rho: E \rightarrow TM$ and a symmetric bilinear form $\langle -, - \rangle$ on the bundle satisfying the following conditions:
\begin{align}
[e^1, [e^2, e^3]_D]_D &= [[e^1, e^2]_D, e^3]_D + [e^2, [e^1, e^3]_D]_D, \notag \\
\rho(e^1)\langle e^2 , e^3 \rangle &= \langle [e^1, e^2]_D, e^3 \rangle + \langle e^2, [e^1, e^3]_D \rangle, \notag \\
\rho(e^1)\langle e^2 , e^3 \rangle  &= \langle e^1 , [e^2, e^3]_D + [e^3, e^2]_D \rangle, \notag
\end{align}
where $e^1$, $e^2$, $e^3\in \Gamma(E)$. The operation $[-,-]_D$ is called \emph{Dorfman bracket} and the map $\rho$ is also called \emph{anchor map}.

Let us discuss some examples of Courant algebroids. The well-known standard Courant algebroid on the \emph{generalized tangent bundle}, $TM\oplus T^* M$, contains the following operations. The Dorfman bracket is given by
\begin{equation}
	[X + \alpha, Y + \beta]_D \equiv [X, Y] + L_X \beta - \iota_Y d \alpha,
\end{equation}
where $X + \alpha,Y + \beta\in\Gamma(TM\oplus T^* M)$. The operation $[X, Y]$ denotes the ordinary Lie bracket on the space of vector fields over $M$. $L_X$ denotes the Lie derivative along the vector field $X$ and $\iota_Y$ the interior product. Finally, $d$ is the de Rham differential over $M$. In general, the Dorfman bracket is not antisymmetric. The antisymmetrization of the Dorfman bracket defines the Courant bracket. The bundle map is the projection to the tangent bundle,
\begin{equation}
	\rho(X + \alpha) \equiv X.
\end{equation}
Finally, the bilinear form on the bundle is given by
\begin{equation}
	\langle X + \alpha, Y + \beta \rangle \equiv \iota_X \beta + \iota_Y \alpha.
\end{equation}
These three operations can be shown to satisfy the conditions of a Courant algebroid.

It is well known, that in general a QP-manifold of degree $2$ is equivalent to a Courant algebroid\cite{Roytenberg99}. The standard Courant algebroid 
can be recovered from a QP-manifold as follows. We take the graded manifold $\cM = T^{*}[2]T[1]M$, where $M$ is a smooth manifold. $[n]$ denotes shift of the fiber degree by $n$, so that the local coordinates are ($x^i$, $q^i$, $p_i$, $\xi_i$) of degree ($0$, $1$, $1$, $2$). Then, we choose the graded symplectic structure as
\begin{equation}
	\omega = \delta x^i \wedge \delta \xi_i + \delta q^i \wedge \delta p_i.
\end{equation}
Finally, the Hamiltonian function is given by
\begin{equation}
	\Theta = \xi_i q^i, \label{StandardCourantTheta}
\end{equation}
which is of degree $3$. More terms in the Hamiltonian function are possible and correspond to twists or deformations of the resulting Courant algebroid. The standard Courant algebroid on the generalized tangent bundle, $TM \oplus T^* M$, can be recovered using \emph{derived brackets}. For this, we introduce the injection map,
\begin{align}
	j: E \oplus TM &\rightarrow \cM \notag \\
	\left(x^i, \partial_i, dx^i, \frac{\partial}{\partial x^i}\right) &\mapsto (x^i, p_i, q^i, \xi_i). \notag
\end{align}
Note that the supergeometry grading is shifted.
Obviously, sections of the generalized tangent bundle correspond to functions of degree $1$ on $\cM$. The Dorfman bracket on $E$ is defined by
\begin{equation}
	[e^1, e^2]_D \equiv - j^*\{\{ j_* e^1, \Theta\}, j_* e^2 \},
\end{equation}
the bundle map via
\begin{equation}
	\rho(e) f \equiv - j^* \{\{j_* e, \Theta\}, j_* f \},
\end{equation}
and the symmetric bilinear form by
 \begin{equation}
	\langle e^1 , e^2 \rangle \equiv j^*\{j_* e^1, j_* e^2 \},
\end{equation}
where $e^1, e^2, e\in\Gamma(E)$ and $f\in C^{\infty}(M)$. In this case, the classical master equation, $\{\Theta, \Theta\} = 0$, is solved trivially and the resulting structure is the untwisted standard Courant algebroid.

A further example is the $H$-twisted standard Courant algebroid. In this case, the operations are defined as follows. The Dorfman bracket is twisted by an $H$-flux term via
\begin{equation}
	[X + \alpha, Y +\beta]_{D,H} \equiv [X,Y] + L_X \beta - \iota_Y d\alpha + \iota_X \iota_Y H,
\end{equation} 
where $X + \alpha, Y +\beta\in\Gamma(E)$ and $H\in \Omega^{3}(M)$ is a closed $3$-form. The symmetric bilinear form as well as the bundle map are not deformed,
\begin{align}
	\langle X + \alpha, Y + \beta \rangle &\equiv \iota_X \beta + \iota_Y \alpha, \notag \\ 
	\rho(X + \alpha) &\equiv X. \notag
\end{align}

The $H$-twisted standard Courant algebroid can be easily reconstructed on a graded symplectic manifold by adding a deformation term to the Hamiltonian function, giving
\begin{equation}
	\Theta_H = \xi_i q^i + \frac{1}{3!}H_{ijk}(x) q^i q^j q^k.
\end{equation}
Now, the classical master equation is not trivially solved, but requires the $H$-flux to be a closed $3$-form $dH = 0$. Using the derived brackets defined above we recover the operations of the $H$-twisted standard Courant algebroid.

Finally, let us give an example of a QP-manifold of degree $2$, which is twisted by a function of degree $2$ and how the twist influences the resulting Courant algebroid. For this, we start again from the untwisted standard Courant algebroid.
We introduce a $2$-form $B$-field such that $dB$ is globally defined\footnote{The $B$-field is understood in the framework of gerbes.}, and 
which in graded manifold language can be expressed by $B = \frac{1}{2} B_{ij}(x) q^i q^j$. 
This function is obviously of degree $2$ and therefore a degree-preserving twist on the QP-manifold is possible. We twist the Hamiltonian function \eqref{StandardCourantTheta} via exponential adjoint action,
\begin{align}
	e^{-\delta_B}\Theta &= \xi_i q^i + \frac{1}{2}\partial_i B_{jk}(x) q^i q^j q^k \notag \\ 
	&= \xi_i q^i + \frac{1}{3!} H_{ijk} q^i q^j q^k,
\end{align}
where now $H = dB$. In this case, the classical master equation is solved trivially, since $dH = d^2 B = 0$. This can be directly seen by investigation of the classical master equation,
\begin{equation}
	\{e^{-\delta_B}\Theta, e^{-\delta_B}\Theta \} = e^{-\delta_B}\{\Theta,\Theta \} = 0, \notag
\end{equation}
due to $\{\Theta,\Theta \} = 0$ for the initial Hamiltonian function. We conclude, that twists of Hamiltonian functions do not induce additional restrictions. The resulting Courant algebroid is the $dB$-twisted standard Courant algebroid. 
In general, other twists are possible and lead to further deformations of the Courant algebroid structure. Instead of classifying all twists, we discuss double field theory and its relation to graded symplectic manifolds in the next section.

\section{Double field theory}

The T-duality group for closed string theory compactified on a $D$-dimensional torus $T^D$ is $O(D,D;\mathbb{Z})$. \emph{Double field theory}\cite{Hull:2009mi,Aldazabal:2013dft} is a manifestly $O(D,D)$-invariant theory. It makes this symmetry manifest by introducing an additional set of variables. In the case of a torus compactification of double field theory, T-duality acts as $O(D,D;\mathbb{Z})$-transformation on the extended set of coordinates. The initial torus is extended to a double torus $T^{2D} = T^D \times \tT^D$. $\tT^D$ denotes the dual torus. We want to call the local coordinates $x^i$ on the initial torus \emph{standard coordinates}, whereas the local coordinates $\tx_i$ on $\tT^D$ \emph{dual coordinates}, associated with the winding numbers of closed strings wrapping the internal circles. Standard and dual coordinates can be rearranged via introduction of generalized coordinates $X^I = (\tx_j, x^i)$. In general, capital indices run over the whole double index range $I=1,\dots, 2D$.

The field content of double field theory is the spacetime metric $g$, the Kalb-Ramond $B$-field and the dilaton $\phi$. All fields depend on the whole set of coordinates $X^I$. To make the $O(D,D)$-action manifest, the $B$-field and metric are combined into a so-called \emph{generalized metric},
\begin{equation}
	\mathcal{H}_{MN} = \begin{pmatrix} g^{ij}  & -g^{ik}B_{kj} \\ B_{ik}g^{kj} & g_{ij} - B_{ik}g^{kl} B_{lj} \end{pmatrix}.
\end{equation}

The group $O(D,D)$ is generated by the following diffeomorphisms, $B$-transformations and $\beta$-transformations,
\begin{equation}
	h_{M}^{\Sp N} = \begin{pmatrix} E^i_{\Sp j} & 0 \\ 0 & E_i^{\Sp j} \end{pmatrix}, \quad 	h_M^{\Sp N} = \begin{pmatrix} \delta^i_{\Sp j} & 0 \\ B_{ij} & \delta_i^{\Sp j} \end{pmatrix}, \quad h_{M}^{\Sp N} = \begin{pmatrix} \delta^i_{\Sp j} & \beta^{ij} \\ 0 & \delta_i^{\Sp j} \end{pmatrix}, \notag
\end{equation}
where $E\in\text{GL}(D)$. $B_{ij}$ and $\beta^{ij}$ are antisymmetric tensors. The generators obey
\begin{equation}
	h_M^{\Sp P} \eta_{PQ} h_N^{\Sp Q} = \eta_{MN}, \quad \eta_{MN} = \begin{pmatrix} 0 & \delta^i_{\Sp j} \\ \delta^{\Sp j}_i & 0 \end{pmatrix}.
\end{equation}
This is the invariance structure of the generalized tangent bundle, $E=TM\oplus T^* M$, discussed above,
\begin{equation}
	\langle X + \alpha, Y + \beta \rangle = \iota_X \beta + \iota_Y \alpha,
\end{equation}
where $X + \alpha, Y + \beta\in\Gamma(E)$, and is closely related to generalized geometry. 

As we discussed above, double field theory doubles the spacetime coordinates in order to make the global symmetry group $O(D,D)$ manifest. In other words, the resulting bundle is the generalized tangent bundle over the double space, $E = T\wM \oplus T^* \wM$, where $\wM = M \times \tM$ is the double space and $\tM$ is the dual to $M$. The gauge transformations are doubled and now consist of standard as well as dual diffeomorphisms and two-form gauge transformations. Using the generalized gauge parameter $\xi^I = (\tlambda_i, \lambda^i)$ and generalized derivatives $\partial_I = (\tpar^i, \partial_i)$, the generalized Dorfman bracket, or \emph{D-derivative}, is defined by\cite{Aldazabal:2013dft}
\begin{equation}
	L_{\xi_1} \xi_2^M \equiv \xi^L_1 \partial_L \xi_{2}^M +  (\partial^M \xi_{1,L} - \partial_L \xi_1^M) \xi_2^L.
\end{equation}
The 
\emph{C-bracket} is defined by antisymmetrization of the D-derivative\cite{Aldazabal:2013dft},
\begin{equation}
	[\xi_1, \xi_2]_C \equiv \frac{1}{2}(L_{\xi_1} \xi_2 - L_{\xi_2} \xi_1).
\end{equation}
The closure of the gauge transformations requires the \emph{strong constraint},
or \emph{section condition},
\begin{equation}
	\eta^{IJ}\partial_I \otimes \partial_J = 0.
\end{equation}
Solving the strong constraint by $\tpar^i = 0$ reduces the D-derivative 
to the Dorfman bracket and C-bracket to the Courant bracket. The resulting frame is the supergravity frame. The so-called \emph{weak constraint} is related to the level matching condition of string theory, which for the massless subsector can be written as
\begin{equation}
	\partial_i \tpar^i \phi = 0,
\end{equation}
where $\phi$ denotes any combination of fields.

T-duality acts in form of the $O(D,D)$-generators on the generalized metric, generalized momentum $P^N = (\tp^i, p_i)$\footnote{We are following the convention in Ref.~\citenum{Aldazabal:2013dft}.} and generalized coordinates,
\begin{equation}
	\mathcal{H}_{MN}(X) \mapsto \mathcal{H}_{PQ}(hX) h_M^{\Sp P} h_N^{\Sp Q}, \quad P^M \mapsto h^M_{\Sp N} P^N, \quad	X^M \mapsto h^M_{\Sp N} X^N. \notag
\end{equation}
It leaves the strong and weak constraints invariant.

Introducing generalized vielbeins $E^A_{\Sp M}$, we can decompose the generalized metric via
\begin{equation}
	\mathcal{H}_{MN} = E^A_{\Sp M} S_{AB} E^B_{\Sp N}, \quad S_{AB} = \begin{pmatrix} \eta^{ab} & 0 \\ 0 & \eta_{ab} \end{pmatrix}. \notag
\end{equation}
Indices $A,B,C\dots$ denote flat indices, whereas $I,J,K\dots$ denote curved indices. We have
\begin{equation}
	\eta_{MN} = E^A_{\Sp M} \eta_{AB} E^B_{\Sp N}, \quad \eta_{AB} = \begin{pmatrix} 0 & \delta_{a}^{\Sp b} \\ \delta^{a}_{\Sp b} & 0 \end{pmatrix}, \notag
\end{equation}
where $\eta_{ab}$ is the $D$-dimensional metric. The transformation behavior of the generalized vielbein under generalized diffeomorphisms is given by
\begin{equation}
	L_\xi E^A_{\Sp M} = \xi^P \partial_P E^A_{\Sp M} + (\partial_M \xi^P - \partial^P \xi_M) E^A_{\Sp P}.
\end{equation}
Under the full $O(D,D)$-group, the generalized vielbein is parametrized by\cite{Hassler:2014cc}
\begin{equation}
	E^A_{\Sp M} = \begin{pmatrix} e_a^{\Sp i} & e_a^{\Sp j}B_{ji} \\ e^a_{\Sp j}\beta^{ji} & e^a_{\Sp i} + e^a_{\Sp j}\beta^{jk}B_{ki} \end{pmatrix}.
\end{equation}
The ordinary spacetime metric is decomposed using standard vielbeins via $g_{ij} = e^a_{\Sp i} \eta_{ab} e^b_{\Sp j}$.
The structure of double field theory can be recovered from a graded symplectic manifold of degree $2$ as follows\cite{DeserSaemann:2016,Deser:2014mxa,Heller:2016u}. We start with the P-manifold ($\hcM = T^*[2] T[1] \hM$, $\omega$), where $\hM = M \times \tM$ denotes the double spacetime. We choose local coordinates ($x^M = (x^i, \tx_i)$, $q^M = (q^i, \tq_i)$, $p_M = (p_i, \tp^i)$, $\xi_M = (\xi_i, \txi^i)$)\footnote{Note that we are following the convention in Ref.~\citenum{Heller:2016u}.} of degree ($0$, $1$, $1$, $2$) and define the graded symplectic structure by
\begin{equation}
	\omega = \delta x^M \wedge \delta \xi_M + \delta q^M \wedge \delta p_M.
\end{equation}
We define a vector field of degree $1$, which does not have to be homological, $Q^2\neq 0$, but satisfies $L_Q \omega = 0$. The resulting structure $(\hcM, \omega, Q)$ is called \emph{pre-QP-manifold}\cite{
Heller:2016u}. $Q$ is still Hamiltonian and therefore we can find a Hamiltonian function $\Theta$. We define the Hamiltonian function by
\begin{equation}
	\Theta_{\text{DFT},0} = \xi_M (q^M + \eta^{MN} p_N). \label{DFTHam}
\end{equation}
The classical master equation of this Hamiltonian function is broken,
\begin{equation}
	Q^2 \sim \{\Theta, \Theta\} \sim \xi_i \txi^i  \neq 0.
\end{equation}
The strong constraint of double field theory can be reconstructed via derived brackets,
\begin{equation}
	\{ \{ f, \{\Theta,\Theta\}\}, g\} = 0,
\end{equation}
where $f,g\in\ChM$. The injection of the generalized double tangent bundle, $E = T\hM \oplus T^*\hM$, to the graded manifold is defined via
\begin{equation}
	\hj: E \oplus T\hM \rightarrow \hcM, \quad \left(x^M, \partial_M, dx^M, \frac{\partial}{\partial x^M}\right) \mapsto (x^M, p_M, q^M, \xi_M). \notag
\end{equation}
The structure of double field theory in terms of the D-bracket can be reconstructed via derived brackets,
\begin{equation}
	[e^1, e^2]_D \equiv - \hj^*\{\{ \hj_* e^1, \Theta_{\text{DFT},0}\}, \hj_* e^2 \},
\end{equation}
where $e^1, e^2\in\Gamma(E)$. This shows that the algebra of double field theory can be recovered by making use of graded symplectic manifolds.

\section{Double field theory and Courant algebroids}

In the previous section, we showed that the Hamiltonian inducing the double field theory algebra does not satisfy the classical master equation. On the other hand, the classical master equation implies the section condition. In other words, solving the section condition reduces the theory to a hypersurface inside the double space and induces a Courant algebroid structure on this hypersurface.

We furthermore reviewed the $H$-twisted standard Courant algebroid and how the $dB$-twisted standard Courant algebroid emerges by twist. We will show in this section, that geometric as well as non-geometric flux freedom can be introduced to the Hamiltonian \eqref{DFTHam} by simple twist. The twists that are needed for this endeavour can be classfied into three categories: $B$-transformation, $\beta$-transformation and diffeomorphism. They correspond to the generators of $O(D,D)$.

The introduction of an $H=dB$ term can be done analogously as described in the sections above. However, the introduction of $f$-flux is a little bit more intricate. For this, we will have to introduce a frame bundle to our graded symplectic manifold. A general frame on the double space $T[1]\hM \oplus T^*[1]\hM$ is denoted by $(q^i, p_i, \tq_i, \tp^i)$. The corresponding flat frame lives in $V[1] \oplus V^*[1] \oplus \tilde{V}[1] \oplus \tilde{V}^*[1]$, where $V = \tilde{V} = \mathbb{R}^D$, and is denoted by $(q^a, p_a, \tq_a, \tp^a)$.

Then, we define twists by vielbein fields in the geometric and non-geometric frame,
\begin{align}
	\exp(\delta_e) \equiv \exp(e^{\Sp i}_a (x) q^a p_i), \quad \exp(\delta_{e^{-1}}) \equiv \exp(e_{\Sp i}^a (x) q^i p_a), \notag \\
	\exp(\delta_\te) \equiv \exp(e^{\Sp i}_a (x) \tp^a \tq_i), \quad \exp(\delta_{\te^{-1}}) \equiv \exp(e_{\Sp i}^a (x) \tp^i \tq_a). \notag
\end{align}
In the same manner, we define $B$- and $\beta$-transformations in the geometric and non-geometric frame
\begin{align}
	\exp(\delta_B) &\equiv \exp\left(\frac{1}{2} B_{ij}q^i q^j\right), \quad \exp(\delta_\beta) \equiv \exp\left(\frac{1}{2} \beta^{ij}p_i p_j\right), \notag \\
	\exp(\delta_\tB) &\equiv \exp\left(\frac{1}{2} B_{ij}\tp^i \tp^j\right), \quad \exp(\delta_\tbeta) \equiv \exp\left(\frac{1}{2} \beta^{ij}\tq_i \tq_j\right). \notag
\end{align}
In turns out that the combinations
\begin{align}
	\exp(\delta_\tV) &\equiv \exp(-\delta_\te)\exp(\delta_{\te^{-1}})\exp(-\delta_\te), \notag \\
\exp(\delta_V) &\equiv \exp(-\delta_e)\exp(\delta_{e^{-1}})\exp(-\delta_e) \notag
\end{align}
produce the correct expressions for the local fluxes.

By twist of the Hamiltonian \eqref{DFTHam} by $B$-, $\beta$- and vielbein transformation in the geometric and non-geometric frame, respectively,
\begin{align}
	\exp(\delta_V)\exp(-\delta_\beta)\exp(-\delta_B)\Theta_{\text{DFT},0}, \notag \\
	\exp(\delta_\tV)\exp(-\delta_\tbeta)\exp(-\delta_\tB)\Theta_{\text{DFT},0}, \notag
\end{align}
we recover the local expressions for $H$-, $F$-, $Q$- and $R$-flux as coefficients of both twisted Hamiltonians\footnote{The local expressions are the same for both twisted Hamiltonians.}\cite{Heller:2016u}
\begin{align}
	H_{abc} &= 3(\nabla_{[a} B_{bc]} + B_{[a|m|}\tpar^m B_{bc]} + \tilde{f}^{mn}_{[a} B_{b|m|} B_{c]n}), \notag \\
	F_{bc}^a &= f^a_{bc} - H_{mns}\beta^{si}e^a_{\Sp i} e_b^{\Sp m} e_c^{\Sp n} + \tpar^a B_{bc} + \tilde{f}^{ad}_b B_{dc} - \tilde{f}_c^{ad}B_{db}, \notag \\
	Q_a^{bc} &= \tilde{f}^{bc}_a +\partial_a\beta^{bc} + f^b_{ad}\beta^{dc} - f^c_{ad}\beta^{db} + H_{isr}\beta^{sh}\beta^{rk}e^{\Sp i}_a e^b_{\Sp h} e^c_{\Sp k} \notag \\
&\qquad + B_{am}\tpar^m \beta^{bc} + \tpar^{[b}B_{ae}\beta^{e|c]} + 2B_{[a|e}\tilde{f}^{be}_{d]}\beta^{dc} - 2B_{[a|e}\tilde{f}^{ce}_{d]}\beta^{db}, \notag \\
	R^{abc} &= 3(\beta^{[a|m|}\partial_m\beta^{bc]} + f^{[a}_{mn}\beta^{b|m|} \beta^{c]n} + \tpar^{[a}\beta^{bc]} - \tilde{f}^{[ab}_d\beta^{|d|c]} \notag \\
	&\qquad  + B_{ln}\tpar^l\beta^{[ab}\beta^{|n|c]} + \tpar^{[a}B_{ed}\beta^{|e|b}\beta^{|d|c]} + \tilde{f}^{[a|e|}_n B_{ed}\beta^{|n|b|}\beta^{|d|c]}) \notag \\
	&\qquad - H_{mns}\beta^{mi}\beta^{nh}\beta^{sk}e^a_{\Sp i}e^b_{\Sp h} e^c_{\Sp k}, \notag
\end{align}
where we defined 
\begin{align}
	H_{mns} &= 3(\partial_{[m}B_{ns]} + B_{[m|l|}\tpar^l B_{ns]}), \notag \\
	f_{ab}^c &= 2e_{[b}^{\Sp i}\partial_i e^c_{\Sp j} e_{a]}^{\Sp j}, \notag \\
	\tilde{f}^{ab}_c &= 2 e^{[a}_{\Sp m} \tpar^m e^{b]}_{\Sp j} e^{\Sp j}_c. \notag
\end{align}
Projection to the supergravity frame or winding frame, respectively, by solving the section condition by $\txi^i = 0$ or $\xi_i = 0$ reduces the twisted pre-QP-manifold to a twisted Courant algebroid. The classical master equation of the projected twisted Hamiltonian then amounts to the Bianchi identities among the fluxes. 

For convenience, let us describe the twisted Courant algebroid in the supergravity frame. After projection via $\txi^i = 0$, the twisted double field theory Hamiltonian reduces to
\begin{align}
	\Theta_{B\beta e} &= e_b^{\Sp i} q^b \xi_i + e^b_{\Sp l} \beta^{lm} p_b \xi_m - e^b_{\Sp l} \beta^{lm}\partial_m e_a^{\Sp j} e^a_{\Sp i} q^i p_j p_b + e_b^{\Sp m} \partial_m e^{\Sp j}_a e^a_{\Sp i} q^i q^b p_j \notag \\
	&\qquad + \frac{1}{3!}H_{abc} q^a q^b q^c + \frac{1}{2} F^a_{bc} p_a q^b q^c + \frac{1}{2} Q_a^{bc} q^a p_b p_c + \frac{1}{3!} R^{abc}p_a p_b p_c, \label{DFTGeo}
\end{align}
where
\begin{align}
	H_{abc} &= 3\nabla_{[a}B_{bc]}, \notag \\
	F^a_{bc} &= f^a_{bc} - H_{mns}\beta^{si} e^a_{\Sp i} e_{b}^{\Sp m} e_{c}^{\Sp n}, \notag \\
	H_{mns} &= 3\partial_{[m}B_{ns]}, \notag \\
	Q_a^{bc} &= \partial_a\beta^{bc} + f_{ad}^b \beta^{dc} - f_{ad}^c \beta^{db} + H_{isr}\beta^{sh}\beta^{rk}e_a^{\Sp i}e^b_{\Sp h} e^c_{\Sp k}, \notag \\
	R^{abc} &= 3(\beta^{[a|m|}\partial_m\beta^{bc]} + f_{mn}^{[a}\beta^{b|m|}\beta^{c]n}) - H_{mns}\beta^{mi}\beta^{nh}\beta^{sk}e^{a}_{\Sp i} e^b_{\Sp h}e^{c}_{\Sp k}. \notag
\end{align}
The classical master equation of \eqref{DFTGeo} amounts to the Bianchi identities
\begin{align}
	e_{[a}^{\Sp m}\partial_{|m|} H_{bcd]} - \frac{3}{2}F_{[ab}^e H_{|e|cd]} &= 0, \notag  \\
	e^{[a}_{\Sp l}\beta^{|lm|}\partial_m R^{bcd]} - \frac{3}{2}Q^{[ab}_e R^{|e|cd]} &= 0, \notag \\
	e^d_{\Sp l} \beta^{ln}\partial_n H_{[abc]} - 3 e_{[a}^{\Sp n}\partial_n F_{bc]}^d - 3H_{e[ab}Q^{ed}_{c]} + 3F^d_{e[a} F^e_{bc]} &= 0, \notag \\
	-2e^{[c}_{\Sp l}\beta^{|ln|}\partial_n F^{d]}_{[ab]} - 2e^{\Sp n}_{[a}\partial_n Q_{b]}^{[cd]} + H_{e[ab]}R^{e[cd]} + Q^{[cd]}_e F^e_{[ab]} + F^{[c}_{e[a}Q^{|e|d]}_{b]} &= 0, \notag \\
	3e^{[b}_{\Sp l}\beta^{|ln|} \partial_n Q^{cd]}_a - e_a^{\Sp n} \partial_n R^{[bcd]} + 3 F_{ea}^{[b} R^{|e|cd]} - 3 Q^{[bc}_e Q^{|e|d]}_a &= 0. \notag
\end{align}
Finally, we can compute the associated twisted Courant algebroid via derived bracket construction. It turns out that the anchor map, fiber metric and Dorfman bracket is given by\cite{Heller:2016u}
\begin{align}
	\rho(X+\alpha)f &= (X + \beta^{\sharp} (\alpha) ) f, \quad \langle X + \alpha, Y + \gamma \rangle = X(\gamma) + Y(\alpha), \notag \\
	[X + \alpha,Y + \gamma]_D &= [X,Y]^\nabla_H + [\alpha,\gamma]_{\beta,H}^\nabla - \iota_\gamma\nabla_\beta X - \iota_Y\nabla\alpha + L_X^\nabla \gamma \notag \\
	&\qquad + L^{\nabla,\beta}_\alpha Y + \iota_Y\iota_X H + \iota_Y\iota_{\beta^\sharp(\alpha)}H + \iota_{\beta^\sharp(\gamma)}\iota_X H \notag \\
	&\qquad - \beta^{\sharp}(\iota_Y\iota_{\beta^\sharp(\alpha)}H) - \beta^\sharp(\iota_{\beta^\sharp(\gamma)}\iota_X H) + \iota_\gamma \iota_\alpha R, \notag 
\end{align}
where $f\in\mathcal{C}^\infty(M)$, $X,Y\in\Gamma(TM)$ and $\alpha,\gamma\in\Gamma(T^*M)$. Furthermore, the operation $\beta^\sharp: T^*M\rightarrow TM$ is locally given by $\beta^\sharp(\gamma) = \beta^{ab}\gamma_a \partial_b$. Above expressions contain covariant versions of the $H$-twisted Lie bracket, $H$-twisted Koszul bracket, Lie derivative and Poisson-Lie derivative defined by
\begin{align}
	[X,Y]^\nabla_H &= [X,Y]^\nabla - \beta^\sharp(\iota_Y \iota_X H), \quad [\alpha,\gamma]^\nabla_{\beta,H} = [\alpha,\gamma]^\nabla_\beta + \iota_{\beta^\sharp(\gamma)}\iota_{\beta^\sharp(\alpha)}H, \notag \\
	L^\nabla_X &= \nabla\iota_X + \iota_X\nabla, \quad\quad\quad\quad\quad\quad\;\;\,\, L_\alpha^{\nabla,\beta} = \nabla_\beta \iota_\alpha + \iota_\alpha \nabla_\beta, \notag
\end{align}
and $\nabla_\beta$ is the covariant Lichnerowicz-Poisson differential, where $f$-flux contributes to the torsion of the connection. 

By construction, ($TM\oplus T^*M$, $\langle-,-\rangle$, $[-,-]_D$) is a Courant algebroid. It lives on the supergravity frame inside the double field theory space and is parameterized by the full set of $O(D,D)$ generators: $B$-field, $\beta$-field and vielbein.

To conclude, in this section we showed that the double space pre-QP-structure can be twisted by various $O(D,D)$-generators inducing local flux degrees of freedom. Carving out a hypersurface in the double space by solving the section condition then reduces the algebra to a twisted Courant algebroid parameterized by the fluxes living in this very frame. An $O(D,D)$-transformation then relates different twisted Courant algebroid on dual hypersurfaces and can be interpreted as solving the section condition in a different way.

\section{Poisson Courant algebroid}

In this section, we discuss the role of the Poisson Courant algebroid as a model for $R$-flux in relation to double field theory. For this, we first give an introduction to the structure as a special Courant algebroid. Then, we provide its reformulation as a QP-manifold of degree $2$. 

Let $E = TM \oplus T^* M$ be the generalized tangent bundle over the smooth manifold $M$. We equip $M$ with a Poisson structure $\pi\in\Gamma(\wedge^2 TM)$ such that the pair $(M, \pi)$ becomes a Poisson manifold. The Poisson condition on $\pi$ can be written 
as $[\pi,\pi]_S = 0$, 
where the bracket $[-,-]_S$ denotes the Schouten bracket on the space of polyvector fields over $M$. $d_\pi = [\pi, -]_S$ denotes the Lichnerowicz-Poisson differential, which satisfies $d_\pi^2 =0$ from $[\pi,\pi]_S = 0$.
Furthermore, we introduce a $3$-vector field $R\in\Gamma(\wedge^3 TM)$, so that $R$ is $d_\pi$-closed, $d_\pi R = [\pi,R]_S = 0$. 
We define a bilinear operation
\begin{equation}
	[X + \alpha, Y + \gamma]_R^{\pi} \equiv [\alpha, \gamma]_\pi + L_{\alpha}^\pi Y - \iota_\gamma d_\pi X - \iota_\alpha \iota_\gamma R,
\end{equation}
where $X + \alpha, Y + \gamma\in\Gamma(E)$. $[\alpha,\gamma]_\pi$ denotes the Koszul bracket, defined by $[\alpha, \gamma]_\pi \equiv L_{\pi^\sharp(\alpha)}\gamma - L_{\pi^\sharp(\gamma)}\alpha - d(\pi(\alpha,\gamma))$. The map $\pi^\sharp:T^* M \rightarrow TM$ is written in local coordinates by $\pi^\sharp(\alpha) = \pi^{ij} \alpha_i \frac{\partial}{\partial x^j}$. Furthermore, we define the inner product on $E$ as usual,
\begin{equation}
	\langle X + \alpha, Y + \gamma \rangle = \iota_X \gamma + \iota_Y \alpha.
\end{equation}
Finally, we define the bundle map, $\rho: E \rightarrow TM$, by
\begin{equation}
	\rho(X+\alpha) = \pi^\sharp(\alpha).
\end{equation}
The resulting structure ($E$, $\langle -, - \rangle$, $[-,-]^\pi_R$, $\rho$) with the conditions $d_\pi R  = 0$ satisfies the Courant algebroid conditions and defines the structure of a \emph{Poisson Courant algebroid with $R$-flux}.

There is a distinct analogy between the Poisson Courant algebroid and the standard Courant algebroid.
Through the introduction of the Poisson bivector field and the associated Poisson cohomology, the roles of the tangent and cotangent spaces in the generalized tangent bundle are exchanged. It therefore describes
\emph{contravariant geometry}. This brings us into the position to introduce a $3$-vector twist of the Dorfman bracket of the Poisson Courant algebroid, in natural analogy to the twist the Dorfman bracket of the standard Courant algebroid by a $3$-form.

The QP-manifold formulation of the Poisson Courant algebroid is readily introduced. We take the same graded manifold, local coordinates and graded symplectic structure as in the standard Courant algebroid case, but the form of the Hamiltonian function is different,
\begin{equation}
	\Theta_\pi = \pi^{ij}(x) \xi_i p_j - \frac{1}{2}\frac{\partial \pi^{jk}}{\partial x^i}(x) q^i p_j p_k + \frac{1}{3!} R^{ijk}(x) p_i p_j p_k.
\end{equation}
The operations of the Poisson Courant algebroid are recovered via derived brackets,
\begin{align}
	[e^1, e^2]_D^\pi &\equiv  j^*\{\{ j_* e^1, \Theta\}, j_* e^2 \}, \notag \\
	\rho(e) f &\equiv j^* \{\{j_* e, \Theta\}, j_* f \}, \notag \\
	\langle e^1 , e^2 \rangle &\equiv j^*\{j_* e^1, j_* e^2 \}, \notag
\end{align}
where $e^1, e^2, e\in\Gamma(E)$ and $j$ denotes the usual injection map. The classical master equation, $\{\Theta_\pi, \Theta_\pi\} = 0$, then induces the conditions $d_\pi^2 = 0$ and $d_\pi R = 0$. The resulting structure is the Poisson Courant algebroid with $R$-flux\cite{Bessho:2015tkk}.

\section{Poisson Courant algebroid and double field theory}

In this section, we discuss various proposals for R-flux geometry.
First, we discuss the role of $R$-trivector freedom and compare it to the $H$-twisted standard Courant algebroid.
Finally, we interpret the Poisson Courant algebroid as a certain projection of double field theory on a $D$-dimensional submanifold of the double space. 

As mentioned in the previous section, the introduction of the Poisson tensor serves as a tool to exchange the roles of tangent and cotangent spaces by the map $\pi^\sharp: T^* M \rightarrow TM$. One notes that for non-degenerate Poisson tensor an isomorphic map between Poisson cohomology on polyvector fields and de Rham cohomology on differential forms can be defined using $\pi^\sharp$. We conclude, that there are symmetries between the Poisson Courant algebroid with $R$-flux and standard Courant algebroid
with $H$-flux in terms of $R \leftrightarrow H$ and $d_\pi \leftrightarrow d$. In the same sense as the standard Courant algebroid
can be $B$-twisted to give a $dB$-twisted Courant algebroid, the Poisson Courant algebroid can be $\beta$-twisted via $\exp(-\frac{1}{2}\beta^{ij}p_i p_j)$ to give the $d_\pi \beta$-twisted Poisson Courant algebroid.

Let us 
discuss how the Poisson Courant algebroid fits into double field theory as a bundle over a special $D$-dimensional submanifold in the $2D$-dimensional double space. The first route, we want to take, is to interpret the Poisson Courant algebroid as a twist of the double field theory Hamiltonian function. 

We start from the untwisted double field theory Hamiltonian function on $T^{*}[2]T[1]\hM$,
\begin{align}
	\Theta_{\text{DFT},0} &= \xi_M (q^M + \eta^{MN}p_N) \notag \\
	&= \xi_i (q^i + \tp^i) + \txi^i (p_i + \tq_i).
\end{align}
Then, we twist with the Poisson tensor $\pi$ via $\exp(\frac{1}{2}\pi^{ij}(x) p_i p_j)$ giving
\begin{equation}
	e^{\delta_\pi}\Theta_{\text{DFT},0} = \xi_i (q^i + \tp^i) + \txi^i (p_i + \tq_i) + \pi^{ij} \xi_i p_j - \frac{1}{2} \frac{\partial \pi^{jk}}{\partial x^i}(q^i + \tp^i)p_j p_k.
\end{equation}
The section condition is deformed,
\begin{equation}
	\txi^i \left( 4\xi_i - \frac{1}{2}\frac{\partial \pi^{jk}}{\partial x^i} p_j p_k\right) = 0.
\end{equation}
Solving the section condition by projection to the supergravity frame ($\txi^i = 0$, $\tq_i = 0$, $\tp^i = 0$), the resulting Hamiltonian function is
\begin{equation}
	e^{\delta_\pi}\Theta_{\text{DFT},0}|_{\tx=0} = \xi_i q^i + \pi^{ij} \xi_i p_j - \frac{1}{2} \frac{\partial \pi^{jk}}{\partial x^i} q^i p_j p_k.
\end{equation}
This is the sum of the untwisted standard Courant algebroid 
and the untwisted Poisson Courant algebroid. Due to 
\begin{equation}
	\left\{\xi_i q^i,\pi^{ij} \xi_i p_j - \frac{1}{2} \frac{\partial \pi^{jk}}{\partial x^i} q^i p_j p_k\right\} = 0, \notag
\end{equation}
the associated nilpotent operators
\begin{align}
	Q_{\text{dR}} &\equiv -\left\{\xi_i q^i,-\right\}, \notag \\
	Q_{\pi} &\equiv -\left\{\pi^{ij} \xi_i p_j - \frac{1}{2} \frac{\partial \pi^{jk}}{\partial x^i} q^i p_j p_k,-\right\}, \notag
\end{align}
define a double complex.

Furthermore, the Poisson Courant algebroid over a $D$-dimensional submanifold $M_\pi \subset \hM$, parametrized by $y^i\in M_\pi$,
\begin{equation}
	\Theta_\pi = \pi^{ij}(y) \eta_i p'_j - \frac{1}{2} \frac{\partial \pi^{jk}}{\partial y^i} q'^i p'_j p'_k,
\end{equation}
is related to the double field theory Hamiltonian over $\hM$ via
\begin{equation}
	\tp^i = q'^i, \quad \tq_i = p'_i, \quad \txi^j = \pi^{ij}(y)\eta_i + \frac{1}{2}\frac{\partial \pi^{jk}}{\partial y^i} q'^i p'_k, \label{PC}
\end{equation}
under the projection to the winding frame ($\xi_i = 0$, $q^i = 0$, $p_i = 0$). Under the assumption that the Poisson tensor is non-degenerate, we find
\begin{equation}
	\tx_i = \int \pi^{-1}_{ij}(y) dy^j.
\end{equation}

We conclude that the Poisson Courant algebroid appears as a solution of the section condition in double field theory.
The Poisson Courant algebroid lives in the double field theory winding space equipped with a Poisson tensor. This explains the appearance of the Poisson connection in \eqref{PC}. The $3$-vector field $R$ of the Poisson Courant algebroid lives in the deformed double field theory winding frame.

One can introduce an $R$-flux freedom by $\beta$-twist of the double field theory Hamiltonian function in the supergravity frame ($\txi_i = 0$, $\tq_i = 0$, $\tp^i = 0$). The resulting flux is $R = \frac{1}{2}[\beta,\beta]_S$ and is complementary to the $R$-flux freedom described by the Poisson Courant algebroid since the solution of the section constraint is orthogonal. In this case, $R$ measures the failure of the $\beta$-induced Poisson bracket, $\{-,-\}_\beta$, to satisfy the Jacobi identity\cite{Blumenhagen:2012Bi}.

\section{Summary and discussion}

In this article, we provided a brief introduction in the vast field of QP-manifolds and Courant algebroids and gave a short review of the ideas of double field theory. After that, we reviewed the construction of the double field theory algebra using pre-QP-manifolds of degree $2$.

The main part concerned the description of how different Courant algebroids emerge as solutions of the double field theory section condition. We computed the twisted Courant algebroid living in the supergravity frame of double field theory. Then, we described the Poisson Courant algebroid as model for $R$-flux and discussed it regarding to the double field theory set-up. We found that the Poisson Courant algebroid with $R$-flux is in analogy with the standard Courant algebroid 
with $H$-flux due to the introduction of the Poisson tensor relating associated Poisson and de Rham cohomologies.

Furthermore, we could derive the Hamiltonian function of the untwisted standard Courant algebroid 
with Poisson Courant algebroid part from a twist of the double field theory Hamiltonian function in the supergravity frame. This Hamiltonian function induces a double complex, since the twist was induced by a Poisson tensor. On the other hand, we could interpret the Poisson Courant algebroid with $R$-flux as a solution of the double field theory section condition in favor of the winding frame. In this case, the winding frame is distorted by the presence of the Poisson tensor and a Poisson connection is induced on the associated bundle. The $3$-vector freedom of the Poisson Courant algebroid then lives in this distorted winding space. 

Since the failure of the double field theory Hamiltonian to induce a Courant algebroid is measured by the section condition, each twisted Courant algebroid will constitute a solution to the section condition in terms of a bundle fibered over a $D$-dimensional submanifold $X\subset\hM$. In this sense, the fluxes $H$, $F$, $Q$ and $R$ constitute different degrees of freedom depending on the Courant algebroid solving the section condition. 

\noindent
\setcounter{equation}{0}
\subsection*{Acknowledgments}
\noindent

The authors thank U.~Carow-Watamura, T.~Kaneko, Y.~Kaneko and the members of the institute for helpful comments and discussions.

M.A.~Heller is supported by Japanese Government (MONBUKAGAKUSHO) Scholarship
and
N.~Ikeda is supported by the research promotion program grant 
at Ritsumeikan University.

\bibliographystyle{ws-procs9x6} 


\end{document}